\documentstyle[12pt,aasms,tighten]{article}
\begin{document}
\title{A non-parametric and scale-independent method for cluster analysis II:
the multivariate case}
\author{ A. Pisani}
\affil{Dipartimento di Astronomia, Universit\`a di Padova, Italy, \\ and \\
International School for Advanced Studies (S.I.S.S.A.), Trieste, Italy.\\
 Author address:\\
 S.I.S.S.A. \\
Via Beirut 2-4, 34013 - Trieste, Italy.\\E-mail: 38028::PISANI}
\newcommand{\be}{\begin{equation}}
\newcommand{\ee}{\end{equation}}
\newcommand{\cp}{ \\ \hspace{1cm} }
\begin{abstract}

 A general method is described for detecting and analysing galaxy
systems. The
multivariate geometrical structure of the sample is studied by using an
extension of the method which we introduced in a previous paper.
 The method is based on an estimate of the probability density underlying a
data sample. The density is estimated by using an iterative and adaptive
kernel estimator. The used kernels have spherical symmetry, however we
describe a method in order to estimate the locally optimal shape of the
kernels.
 We use the results of the geometrical structure analysis
 in order to study the effects that is has on the cluster parameter estimate.
 This suggests a possible way to distinguish between structure and
substructure within a sample.

  The method is tested by using  simulated numerical models and applied to two
 galaxy samples taken from the literature.
 The results obtained for the Coma cluster suggest a
 core-halo structure formed by a large number of geometrically
independent systems. A different conclusion is suggested by
the results for the Cancer cluster indicating the presence of at least
two independent structures both containing
 substructure.

 The dynamical consequences of the results obtained from the geometrical
 analysis will be described in a later paper.

  Further applications of the method are suggested and are currently in
progress.

\end{abstract}


\section{Introduction}

  Cluster analysis includes a rather rich list of methods
that mathematical statistics offers in order to group sample data into simpler
subunits (e.g. Murtagh \& Heck, 1987;
Pisani, 1994, hereafter paper I, and references therein)
 Within the astronomical and astrophysical context,
cluster analysis can be applied to galaxy samples spanning large scale volumes
in order to detect systems
generally termed groups and clusters of galaxies. On the other hand, by
studying
small-scale volumes containing rich concentrations of galaxies it is possible
 to analyse substructures within
systems (i.e. clusters) of galaxies. As  was described in
paper I,
 the methods of cluster analysis can be broadly divided into
two families: parametric and non-parametric methods. In astronomy,
non-parametric methods are hardly ever used. On the other hand there is a very
rich literature of examples of applications of parametric methods. These
 methods generally work in the following way. First it is necessary not only to
define what a system is operationally, but also to give a critical or fiducial
value of a particular parameter that can be estimated for each system
such as, for
example, the local luminosity density of galaxy groups as
in Tully (1980, 1987) or the separation in velocity and angular coordinates
 as in Huchra \& Geller (1982, and more recently
Ramella, Geller \& Huchra 1989).
 Then the
parametric method detects only those systems whose parameter is larger than a
given threshold. It is not difficult to understand that within this framework
a lot of parameters can be taken and an arbitrary choice of the critical
parameter can lead to an author-dependent catalogue of systems. This
family of methods is well suited to detecting structures that are well known in
advance, but they risk  introducing strong biases if the features of the
systems to be detected are not well known as  is the case for galaxy
clustering in astronomy. These facts are the main cause of the
disagreement that exists among the results obtained by several authors
in  both large scale cluster  analysis and small scale substructure
analysis (see e.g. paper I).

 In order to overcome these problems, we propose here a method that
satisfies the
following requirements: a) it is non-parametric, in the sense that no
particular
assumption needs to be made concerning the number, shape or
any parameter or feature of the systems
that are supposed to be detected;
 b) it gives an estimate of the statistical
significance according to which the detected systems can be distinguished from
noise fluctuation; c) it gives an estimate of the probability that a given
galaxy is a member of a given system, for all galaxies and all systems; this
 information  can be used to detect interlopers within systems; d)
 it can be used to infer
considerations about the dynamics of the systems within the data sample,
at least in the case that galaxy catalogues are being considered.  Hence
the structure
 of the method consists of a geometrical analysis whose results are
the basis of the subsequent dynamical analysis. We must stress that  the
geometrical analysis of the clustering structure of galaxy samples assumes
that the
velocity of a galaxy is a pure radial coordinate. This assumption
is trivially false within a gravitationally bound system; however without
assuming this the
geometrical analysis cannot be done. We will discuss the effects of this
assumption in the dynamical part of the
analysis that will be presented in a future paper.

 The main mathematical details of the univariate version of the
method we propose are described in
paper I.
 In the present paper
we  consider the multivariate extension of the method.
 The geometrical
clustering analysis is based on the estimate of the probability density
function underlying the data sample. The extension of this estimate can
profitably be used to study cosmological density fields. This and other
applications of the method are currently in progress.

 The present paper is structured as follows. In \S 2 the extension of the
univariate non-parametric method described in paper I to the general
multivariate case is described. In \S 3 we consider the problem of estimating
the density of a highly elongated system by using spherical kernels.
 In \S 4 a correction is presented and tested for taking into account the
local optimal shape of kernels in the density estimate.
 We describe
in \S 5 a modified version of the Kittler (1976) mapping. This
mapping is basically a visualization tool, but it can be used profitably
for obtaining information concerning the effects that the detected structure
has on some relevant cluster parameters.
In \S 6 we consider a possible way to distinguish between the presence of
structure and substructure within a given sample.
 In \S 7 we describe the general
structure of
the method of clustering analysis we propose.
In \S 8 and 9 we report the results obtained from
the analysis of two well known galaxy clusters. In this paper we consider only
galaxy samples spanning small scale volumes.
The analysis of galaxy
samples spanning large volumes will be treated in a future paper. Finally,
\S 10 summarizes the main points and the results obtained.

\section{The extension to the multivariate case}

\subsection{The estimate of the probability density and
the  definition of cluster}

   In principle, the extension of DEDICA (paper I) to the case of a number of
 dimensions $d$ larger than one is trivial. In fact, the fundamental relations
used in paper I hold in $d$ dimensional spaces in the following form.

  With reference to the definitions given in \S 3.1 of paper I, we suppose
that we have a sample $D_{N}$
of $N$ points with position vectors $\vec{r}_{i} \in \Re^{d}$,
$(i=1,\ldots,N)$.
 If $\sigma_{i}$ is the kernel width of the $i^{th}$ point, the adaptive
kernel density
 estimate $f_{ka}(\vec{r})$ is given by:
\be
f_{ka}(\vec{r}) = \frac{1}{N} \sum_{i=1}^{N} K(\vec{r}_{i},\sigma_{i};
\vec{r})
\label{fk}
\ee
where $\vec{r}$ are position vectors and
we consider the Gaussian kernel:
\be
K(\vec{r}_{i},\sigma_{i};\vec{r}) = \frac{1}{(\sqrt{2\pi} \sigma_{i})^{d}}
\exp\left[-\frac{1}{2} \frac{|\vec{r}_{i}-\vec{r}|^{2}}{\sigma_{i}^{2}}
\right].
\label{krsr}
\ee

 In order to completely specify the density estimate $f_{ka}(\vec{r})$
we have to describe the procedure for estimating the kernel widths
$\sigma_{i}$.
 We adopt here the same procedure described in paper I: the local minimization
of the integrated square error:
\be
ISE[f] \equiv \int_{\Re} [F(\vec{r}) - f(\vec{r})]^{2} d\vec{r}
\ee
where $F(\vec{r})$ is the "true" probability density field. The use of a
kernel designed to minimize the integrated square error in order to study the
modal points of a distribution  was addressed in paper I where we have obtained
encouraging results for the univariate case.
It is easy to
note that:
\be
ISE[f] = \int_{\Re} F^{2}(\vec{r}) d\vec{r} + \int_{\Re} f^{2}
(\vec{r}) d\vec{r} - 2 \int_{\Re} F(\vec{r}) f(\vec{r}) d\vec{r}
\ee
and hence the minimization of $ISE[f]$ is equivalent to the minimization
of the cross validation $M(f)$ defined by:
\be
M(f) \equiv \int_{\Re} f^{2}(\vec{r}) d\vec{r} - 2 \int_{\Re}
F(\vec{r}) f(\vec{r}) d\vec{r}.
\ee
It is possible to show that, under mild assumptions, $M(f)$ can be
expressed as a function of the sample data positions $\vec{r}_{i}$ and kernel
widths $\sigma_{i}$ (see paper I and references therein).

 The iterative method which we propose for selecting the kernel widths
of eq.(\ref{fk}) is the following:

\begin{enumerate}
\item
set the first guess of the size
$
\sigma_{n = 0} = 4 \sigma_{t},
$
where $\sigma_{t}$ is the rule-of-thumb prescription for $\sigma$,
see eq.~(\ref{sigt})
(note that there is no particular reason to choose the factor four in the
previous equation, the aim is to start with a large value of $\sigma_{0}$),
\item
take the value at the $n^{th}$ iteration to be
$\sigma_{n} = \sigma_{n-1}/2$
\item
 take the adaptive kernel estimate $f_{ka}^{(n)}(\vec{r})$ defined by
\begin{itemize}
\item $f_{p}(\vec{r}) = $ fixed
 kernel estimate with kernel sizes $\sigma_{i}=\sigma_{n} \; \forall i$
\item sensitivity  $\alpha=1/2$ (see paper I for further details)
\end{itemize}
\item compute the value of $M(f_{ka}^{(n)})$
\item finally select the optimal $\sigma_{n}$ by minimizing the value of
$M(f_{ka}^{(n)})$ and take the corresponding $f_{ka}(\vec{r})$ as the
optimal density estimate.
\end{enumerate}
 By examining the tests we have done,
 it turned out that less than ten steps
$(n \leq 10)$ are enough to reach the optimal size, in agreement with
the behaviour
of this procedure found for the univariate case examined in paper I.

The
 rule-of-thumb estimate of the kernel size $\sigma_{t}$ is given by:
\be
\sigma_{t} = A(K) N^{-\frac{1}{d+4}} \sqrt{\frac{1}{d} \sum_{l=1}^{d}
 s_{l l}^{2}}
\label{sigt}
\ee
where $s_{l l}$ is the standard deviation of the $l^{th}$ coordinate
 ($l=1,\ldots,d$) of the sample data. The constant $A(K)$ depends
only on the kernel and for the Gaussian case can be taken as $0.96$
(Silverman 1986).

 To complete the extension of the density estimate to the multivariate case
it is sufficient to give the expression for the cross validation:
\be
M(f) = \frac{1}{N^{2}} \sum_{i=1}^{N} \sum_{j=1}^{N} K^{(2)}(
|\vec{r}_{i}-\vec{r}|;\sigma_{i},\sigma_{j})- \frac{2}{N(N-1)}
\sum_{i=1}^{N} \sum_{j \ne i} \frac{1}{\sigma_{j}^{d}}
k\left(\frac{|\vec{r}_{i}
-\vec{r}_{j}|}{\sigma_{j}}\right)
\label{cv}
\ee
where
\be
k(t) = \frac{1}{(2 \pi)^{d/2}} \exp\left( -\frac{1}{2} t^{2}
\right)
\label{k}
\ee
and
\be
K^{(2)}(t;\sigma_{i},\sigma_{j}) = \frac{1}{[2 \pi
(\sigma_{i}^{2}+\sigma_{j}^{2}
)]^{d/2}} \exp\left( -\frac{1}{2} \frac{t^{2}}{\sigma_{i}^{2} + \sigma_{j}^{2}}
\right).
\ee
 We must stress that we adopt here the Gaussian kernel (eq.[~\ref{krsr}] and
[\ref{k}]) because it is differentiable in all its domain and this is a crucial
feature in order to correctly use eq.(\ref{rm}) in order to find local maxima
in the estimated probability density field eq.~(\ref{fk}). However the
 Gaussian  kernel is not the most efficient choice in order to minimize the
$ISE[f]$. In fact the Epanechnikov kernel (Epanechnikov 1969)
performs better, however Silverman (1986)  shows that the values of
the $ISE[f]$ obtained for these kernels are very similar. The choice of
the radial dependence (exponential as for the Gaussian kernel or quadratic as
for the Epanechnikov kernel)
of the kernel $k(t)$ is not crucial in this sense,
 as far as $k(t)$ satisfies the normalization and convergence conditions
described in \S 3.1 of paper I (see also Silverman 1986). What is crucial for
the $ISE[f]$ is the width of the kernels.

 Finally we assume, as in paper I, that a peak in the density function
indicates the presence of a cluster in the data.
The  position vectors
of the peaks
of the probability density estimate are defined as in paper I
by the solution of the iterative equation:
\be
\vec{r}_{m+1} = \vec{r}_{m} +
 a_{d} \frac{\vec{\nabla} f(\vec{r}_{m})}
{f(\vec{r}_{m})}
\label{rm}
\ee
with
\be
a_{d} = \frac{d}{ \sum_{i=1}^{N} \left[ \frac{\vec{\nabla}f(\vec{r}_{i})
}{f(\vec{r}_{i})} \right]^{2}}
\ee
(see Fukunaga \& Hoestetler, 1975).

Since the limit $\vec{\varrho}$
of the sequence in eq.~(\ref{rm}) generally depends on the initial
value of the position vector $\vec{r}_{m=1}$:

\be
\lim_{m\rightarrow +\infty} \vec{r}_{m} = \vec{\varrho}(\vec{r}_{m=1})
\ee
we can label each point in the sample $D_{N}$ by the limit
 $\vec{\varrho}_{i}=\vec{\varrho}(\vec{r}_{m=1}=\vec{r}_{i})$
reached by the sequence
in eq.~(\ref{rm}) starting from the position of the $i^{th}$ point.
 Then we can define
the cluster $C_{\mu}$ as the set of all the points of $D_{N}$ having the same
label $\vec{r}_{\mu}$ and containing $n_{\mu}>1$ points:
\be
C_{\mu} = \{\vec{r}_{i}: \vec{\varrho}_{i}= \vec{\varrho}_{\mu};
n_{\mu} >1 \}
\ee
with $i=1,\ldots,N$ and
$\mu=1,\ldots,\nu$ assuming that $\nu$ is the number of systems present
within $D_{N}$.

 The set of labels $\varrho$ that are associated with only single
points form the
population of isolated points $C_{0}$, hence:
\be
C_{0} = D_{N} - \bigcup_{\mu=1}^{\nu} C_{\mu}.
\ee

  Equations (1) to (10) completely define the  extension
of the density estimation to
the multivariate case. The scale independence of the present density estimate
can be proved following the same argument used in paper I.
The ability of the density estimate  to describe highly elongated
systems
by using a centrally symmetric kernel  is
tested in \S 3 and 4.

\subsection{The quantitative description of the clustering pattern}

 The statistical significance of the $\mu^{th}$ cluster
$\cal{S}_{\mu}$ ($\mu=1,\ldots,\nu$)
is defined as in paper I. In particular it is based on the assumption that
the presence of the $\mu^{th}$ cluster produces a higher value of the local
density and hence of the sample likelihood $L_{N} = \prod_{i} [ \sum_{\mu=0}
^{\nu} F_{\mu}(\vec{r}_{i}) ]$ relative to the value $L_{\mu}$:
\be
L(\mu) = \prod_{i}\left[
f_{ka}(\vec{r}_{i})-F_{\mu}(\vec{r}_{i}) +
\frac{1}{N}\sum_{j\in C_{\mu}} K(\vec{r}_{j},\sigma_{0};\vec{r}_{i})\right]
\label{lmu}
\ee
that one would have if the members of the $\mu^{th}$ cluster were all isolated
and hence belonging to the background component. A large value of the
ratio $L_{N}/L_{\mu}$ characterizes the most prominent clusters. By using the
likelihood ratio test it is possible to estimate the cluster significance
${\cal S}_{\mu}$
 (Materne 1979 and references therein).
The contribution to the local density $f_{ka}(\vec{r})$ due to the $\mu^{th}$
cluster is:
\be
F_{\mu}(\vec{r}) = \frac{1}{N} \sum_{j \in C_{\mu}} K(\vec{r}_{j},\sigma_{j};
\vec{r})
\label{fmu}
\ee
(see also paper I),
and the kernel size of the background component is estimated
by:
\be
\sigma_{0} = max_{i}\{ \sigma_{i} \}
\label{sig0}
\ee
which is only slightly different from the estimate given in paper I, where
$\sigma_{0}$ was defined as the average width of the kernels associated to
the isolated data points.
 The set
of the $n_{\mu}$ members of the $\mu^{th}$ cluster is indicated by $C_{\mu}$.
The function $F_{\mu}(\vec{r})$ is defined as in paper I.

  Another modification compared to paper I concerns the estimate of the
membership probabilities $P(i \in \mu)$ that the $i^{th}$ point is member of
the $\mu^{th}$ cluster.
 As the dimensionality $d$ of the sample data increases, the
discreteness effect due to the fact that we observe points and not a continuous
function causes the $i^{th}$ kernel $K(\vec{r}_{i},\sigma_{i};\vec{r}_{i})$
to give a larger and larger contribution to $f_{ka}
(\vec{r}_{i})$ as $d$
grows. This fact however does not necessarily mean that the  "real"
probability of isolation of the $i^{th}$ point is larger for larger $d$.
 A general multivariate estimate of $P(i \in \mu)$ can be
obtained by noting that the
isolated points have:
\be
f_{ka}(\vec{r}_{i})
\sim \frac{1}{N} K(\vec{r}_{i},\sigma_{0};\vec{r}_{i}).
\ee
 In eq.~(\ref{lmu})
we estimate that the background density field is
given by:
\be
 F_{0}(\vec{r}) = \frac{1}{N} K(\vec{r},\sigma_{0};\vec{r}).
\ee
which supports eqs.~(\ref{lmu}) and (\ref{sig0}).

 This suggests to define the probability that a given point is isolated, and
hence due to the background component, as:
\be
 P(i \in 0) = \frac{ \frac{1}{N} K(\vec{r}_{i},\sigma_{0};\vec{r}_{i})}
 {f_{ka}(\vec{r}_{i})}.
\label{pin0}
\ee
 Numerical tests have shown that this expression is a more general estimate
of the isolation probability than the definition adopted in paper I.
However the two definitions are quite similar for $d=1$.  The difference
between the definition adopted here and the one given in paper I is due to
their
different choice of the $\sigma_{0}$ value.

The probability that the $i^{th}$ point is not isolated can be estimated as:
\be
P_{i} = P(i \notin 0) = 1 - P(i \in 0) = \sum_{\mu=1}^{\nu} P(i \in \mu),
\ee
in order to satisfy the normalization constraint:
 $\sum_{\mu=0}^{\nu} P(i \in \mu) =1$ (see Materne, 1979).
This definition is also in agreement
with the prescriptions of the Neyman-Pearson theory of tests (see e.g.
Ledermann 1984, vol. VI, pag. 278).

Moreover we assume that:
\be
P(i \in \mu) \propto F_{\mu}(\vec{r}_{i}).
\ee
 We adopt the following
definition:
\be
P(i \in \mu) = P_{i} \frac{F_{\mu}(\vec{r}_{i})}
{f_{ka}(\vec{r}_{i})}
\label{pinmu}
\ee
in other words the probability that a point is a member of a structure is
distributed among
 the clusters according to the contribution to the total
density due to each cluster.

  The given definitions of $P(i \in 0)$ (eq.[\ref{pin0}])
and $P( i \in \mu)$  (eq.[\ref{pinmu}]) are an
extension to the kernel density estimate of the similar quantities defined in
the general case by Materne (1979). Our definitions inherit the features of
asymptotical optimality of the kernel density estimator and hence are well
suited as a non-parametric test to efficiently
recognize and eventually remove background
fluctuations as numerical tests have indicated.
The efficiency of $P( i \in 0)$ and $P_{i}$
 estimates in order to detect the background
component shown in the example discussed in \S 4.
 Recently Ramella et al.
(1995) have used our prescription for $P( i \in 0)$ and $P( i \in \mu)$ in
order to identify outliers in groups of galaxies. They have also applied
other statistical methods and dynamical simulations.
 Their results support our prescription.

\subsection{A measure of the overlapping among clusters}

  It is often useful to have a measure of how close the clusters are to each
other.
 We want here to propose an estimate of the overlapping measure for the
multivariate case that is non-parametric and not only pairwise.
 To this aim we introduce the following
definitions. Suppose that we consider the $\mu^{th}$ cluster $C_{\mu}$
containing $n_{\mu}$ members.
 For
each member $i_{\mu} \in C_{\mu}$ we can estimate $P(i_{\mu} \in 0)$ and
$P(i_{\mu} \in \lambda)$, with $\lambda=1,\ldots,\nu$. The interlopers
(or outliers) are defined as the members that satisfy the following relation:
\be
P(i_{\mu} \in 0) \geq P(i_{\mu} \in \mu)
\ee
in other words they are more probably isolated points than cluster members.

  In the case that for at least one cluster $\lambda \neq \mu$ we have
\be
P(i_{\mu} \in \lambda ) \geq P(i_{\mu} \in 0)
\label{pimuinl}
\ee
for some $i_{\mu} \in C_{\mu}$,
then there are some members of the $\mu^{th}$ cluster that
could have been classified as members of the $\lambda^{th}$ cluster. In
other words for some members of the $\mu^{th}$ cluster the probability
$P_{i_{\mu}}$ to be part of a structure is not entirely due to the presence of
only one cluster. In this case we can say that the $\mu^{th}$ and the
$\lambda^{th}$ cluster are in  contact: they somehow  share some
members, they are not completely disjoint. We express this relation by the
notation:
\be
\mu     \rightarrow \lambda.
\ee

  It is easy to see that the $\mu^{th}$
 cluster can be in contact with several clusters.
The  strength ${\cal L}(\mu,\lambda)$
of the contact can be measured by
\be
{\cal L}(\mu,\lambda) = \sum_{i_{\mu} \in C_{\mu}} P( i_{\mu} \in \lambda) +
\sum_{i_{\lambda} \in C_{\lambda}} P( i_{\lambda} \in \mu).
\label{conta}
\ee
  We can also see that the
contact may be a non-symmetric property of clusters. In other words, if the
$\mu^{th}$ cluster is in contact with the $\lambda^{th}$ cluster it is not
necessarily true that the $\lambda^{th}$ cluster is in contact with the
$\mu^{th}$ cluster. This is formally due to the fact that
 eq.(~\ref{pimuinl}) is not symmetric in $\lambda$ and $\mu$.
Asymmetric contact may occur between a significant and
compact cluster and a non significant loose cluster that happens to fall close
enough to the first cluster.

  In the case of symmetric contact we write:
\be
\lambda \leftrightarrow \mu.
\ee

 The above definitions are crucial for examining substructured and elongated
clusters.

\section{The effect of anisotropy}

 As introduced in the previous section, we want to examine here the
following question: how reliable is the estimate of the density of
an elongated system obtained by using centrally symmetric kernels?

 The analytical answer to this question comes from a theorem
(Fukunaga 1972, Silverman 1986)
that proves the following statement.
 Suppose we have a multidimensional
and unimodal system with covariance matrix ${\cal A}$. If we adopt
a kernel estimate of the probability density,
 then
the $ISE[f]$ is reduced if we change the metric of our space. Instead of
the usual euclidean metric, represented by the unit matrix ${\cal I}$,
it is more convenient to use the metric represented by the inverse of the
 covariance matrix ${\cal A}^{-1}$. This result holds independently of the
kind of kernel adopted (not necessarily Gaussian, Silverman 1986).
 If ${\cal A}$ is not the identity matrix ${\cal I}$, we can change the metric
of our space by: $\vec{y} = {\cal W} \vec{r}$.
 Here
${\cal W}$ is the $d \times d$ matrix that describes the
 transformation taking the
covariance matrix of the data into the identity matrix. This transformation
is also called whitening (Fukunaga 1972). The statement of the theorem is
equivalent to saying that in the whitened space the centrally symmetric
kernel (see eq.~[\ref{k}]) has the optimal\footnote{In general, the
optimality of the kernel estimate is referred to asymptotic conditions, i.e.
in the limit $N \;  \rightarrow \; +\infty$ (see paper I and references
 therein).}
 shape, in the sense that it gives a smaller
value of the $ISE[f]$ relative to the original metric (or shape). The
expression for the kernel density estimate taking into account the change in
the
metric is indicated in the next section (eq.~[\ref{fw}]).

  By using $N_{s}=1000$
Monte Carlo simulations of a model density field $F(x,y)$ with $N_{p}=500$
points
we have tested the performance of the
density estimator described on \S 2 for systems with rather different
anisotropies. The goodness of the density estimate is measured by:
\be
\Delta = \frac{1}{100 \times 100} \sum_{i=1}^{100} \sum_{j=1}^{100}
 [F(x_{i},y_{j}) - f_{ka}(x_{i},y_{j})]^{2}
\label{delta}
\ee
for each simulation.
In our tests we have taken for the model field $F(x,y)$ a
 bivariate Gaussian with several values of the axial ratio $a/b$.
 The relevant statistical parameters of $\Delta$ are
reported in table~\ref{tab_delta} for several values of $a/b$.
 In each simulation the
model field has unit variance along the major axis, while the kernels are
centrally symmetric.
 The results of the simulations are shown in Figs.~1 and 2
(mean values are plotted).
 In Fig.~1 we show the results obtained in the case of a centrally symmetric
model density field $F(x,y)$ and by using a Gaussian kernel with the same
central symmetry.
In Fig.~2 we show the case of a model density field with
an elliptical shape with axis ratio $a/b=0.1$, while the used kernels are
Gaussian with central symmetry and hence do not have the same shape of the
model field (they are not optimal).
 It
appears that in the case where the kernel has the locally optimal shape
(Fig.~1) the estimate is rather accurate, confirming the quoted theorem.
In the case of a kernel which does not have the optimal shape (Fig.~2)
rather rounder
isoplethes are obtained. However the estimate of the density is still good.
The quantity that is more sensitive to the non-optimality of the kernel
 shape is the gradient of the density. In the next section we show that it
can fluctuate quite widely as will be illustrated in the example given.
 We do not report the results of the test in the case that both the data and
the
kernels have the same value of axis ratio $a/b<1$ since in that case the
whitening procedure make this test equivalent to what obtained in the case
$a/b=1$.
  We note that the value of $N_{p}$ used in our simulations is rather high
relative to the size of usual observational samples (see \S~8 and 9).
The aim of these simulations is to illustrate the two different
behaviours of the kernel
density estimator when the kernel shape is optimal and when it is not,
in the asymptotic limit.
We also note that for asymptotic conditions we assume we have
no problems in the estimate of the covariance matrix, contrary to what is
likely in real situations.
In the next section we analyse the problems that
arise when dealing with smaller anisotropic systems imbedded in an isotropic
background.

\section{An estimate of the locally optimal metric}

  The message contained in the theorem of the previous section is that
in the general multimodal case, each sample point should be associated
with a matrix which
defines the  locally optimal (in the sense discussed in \S~2.1 and \S~3,
for further details see Fukunaga, 1972)
metric, estimated from the covariance matrix
of the system to which the point belongs. Unfortunately we do not know this
information in advance.
In order to save the non-parametric nature of the
present method, we should avoid all a priori guesses concerning the
number of systems present in our sample and the membership of each
point.

  We propose a way out of this problem based on  very simple remarks. Let us
consider the following example in order to clarify the whole procedure.
We consider for the moment only $d=2$ samples because of their easy geometrical
visualization. In Fig.~3 a $d=2$ sample is shown. It contains $100$ points
obtained by a Monte Carlo simulation of an unimodal system with a bivariate
Gaussian profile, axis ratio $a/b=0.1$ and position angle $\theta =
 45^{\circ}$. Superimposed on this system there is a population of $100$
points randomly distributed. A straightforward application of the $d=2$ version
of DEDICA described in \S~2 indicates the presence of $\nu=26$ clusters and
$n_{0}=0$ isolated points (see Fig.~4).
 The density estimate
$f_{ka}(x,y)$ is shown in
Fig.~5.
 In table~\ref{mc_ds_1} we report the richness and significance
of each cluster. Our experience and a conservative point of view suggests  we
consider a cluster as real if its
 significance is larger
than $0.995$ in
the case of a sample without isolated points, while a cutoff of $0.99$ may be
adopted for samples containing isolated points. This is caused by the fact
that in a sample without isolated points the value of $\sigma_{0}$
(eq.~[\ref{sig0}])
may be underestimated.  As a general empiric
rule (see Materne 1979; Ledermann 1984
vol. VI pag. 278-282) the value 0.99 can be assumed.

  From an inspection of Fig.~4 and 5  it appears clear that the disagreement
between the central symmetry of the kernel used and the locally optimal  metric
has caused the elliptical system to be artificially fragmented into several
subsystems aligned along the major axis of the simulated model profile. We can
also notice that the spurious subsystems are quite close to one another and
it is easy to guess that the density estimates of these subsystems should show
a certain mutual contact (see also Fig.~6). 
By merging the systems that are in mutual contact we
may be able to recover the true elongated system and hence the locally
optimal metric. Spurious contacts due to non-significant clusters and/or
isolated points must be avoided in order to prevent biases in the estimate of
the covariance matrix.

   We adopt the following procedure:
\begin{enumerate}
\item take the first estimate of the clustering structure of the sample by
using centrally symmetric kernels, hence obtaining a catalogue of clusters and
isolated points,
\item remove from the catalogue all the non-significant clusters and isolated
points,
\item check if the remaining systems are in contact and merge all the
 clusters that show mutual contact, thus obtaining a new catalogue of
clusters,
\item consider the new clusters resulting from merger of previous distinct
clusters, check if these clusters are unimodal by applying DEDICA to the
whitened coordinates of the cluster members. If the new cluster
is unimodal, estimate the covariance matrix of the cluster and hence the
locally optimal metric. Otherwise, keep the previous local metric,
\item consider the new estimate of the density by taking into
 account the new local metric determined in the previous step and check if
the new density estimate causes different clusters to be in contact; if
so, return to previous step,
\item if no further merging occurs, then the locally optimal metric of each
point can be estimated from the covariance matrices of the merged clusters
${\cal A}_{i} \neq {\cal I}$, while for the non-merged and non-significant
clusters ${\cal A}_{i}$ is assumed to be the identity ${\cal I}$. The final
density estimate is:
\be
f_{kaw}(\vec{r}) = \frac{1}{N} \sum_{i} \frac{
(det\{ {\cal A}_{i} \})^{-1/2} }{(\sqrt{2 \pi} \sigma_{i})^{d}}
\exp\left[ - \frac{1}{2} \frac{(\vec{r}-\vec{r}_{i})^{T} {\cal A}_{i}^{-1}
(\vec{r}-\vec{r}_{i})}{\sigma_{i}^{2}} \right].
\label{fw}
\ee
where $w$ signifies whitened.

 We can obtain the final classification by using
the peak determination equation~(\ref{rm}) applied to
$f_{kaw}(\vec{r})$.
\end{enumerate}
 In general we have found that this procedure converges quite rapidly (only one
iteration !) to the locally optimal metric.
Let us now see how it works in the trial sample described above.
 By inspection of  table~\ref{mc_ds_1},
it can be seen that the clusters 1, 2,
8, 9, 12, 14 and 17 have significance larger than 0.995 while the cluster
number
19 has a significance of $0.991$. If we include the cluster 19 in the set of
significant clusters, adopting the weak limit of the significance, we obtain
the results reported in table~\ref{ct_1}
 after the contact test: all the clusters are in
mutual contact, hence they are probably part of only one system that may be
 elongated and/or substructured. By performing the test of modality, we can see
that the merged cluster obtained at the step 4 of the above procedure is not
unimodal. Hence the list of clusters that we have merged is either a collection
of actually different clusters close to one another, or we have included in
the list of significant clusters some spurious systems that have biased the
estimate of the locally optimal metric.
To clarify this point we examine the result of the contact test reported
in table~\ref{ct_1}. It can be seen that cluster 19 shows a quite different
behaviour from all the others. It has only asymmetric contacts with
other clusters and it shows contact strength nearly an order of
magnitude weaker than for the other clusters. Finally it has the
smallest value of the significance $S_{\mu}$ and it is characterized by a low
value of $P(i \notin 0)$. The contours plotted in Fig.~7 also show that the
cluster $\mu=19$ should be considered as detached from the other significant
clusters.
 These remarks suggest that
we  re-perform the merging after having removed the cluster $\mu=19$. In
this case we find that all of the clusters considered merge to form a
unique unimodal structure.
The estimate of the local metric is a rather good approximation of the
simulated
model.   In Fig.~8 we show
the plot of the density $f_{kaw}(x,y)$ taking into account the locally
 optical metric. Eventually we can apply the peak identification algorithm
(eq.~[\ref{rm}])
 to check if the membership of the non-merged part of the catalogue,
discarded in step 2,
changes or not. In general this could produce only minor changes in the
membership assignments.

We have preferred to apply the above method, that will whiten the data
sample only in the case of a positive outcome of the contact test,
instead
 of an unconditioned
whitening of the data sample because it analyses more carefully the data
structure.
  We have applied the quoted procedure to several multimodal $d=2$ samples and
also to rich and poor $d=3$ samples always obtaining good results. Hence the
reliability of the procedure which we are proposing is well confirmed.

\section{The Kittler mapping}

        Here, we introduce a useful tool for representing the structure
of multivariate data sets.
We can roughly say that the aim of this tool is to construct a path through
the data set that takes as many as possible data points near a particular
mode (peak), in the estimated probability density $\hat{f}$,
 before moving to another
nearby mode. At each step in the sequence the next point is selected within
"nearby" data points so that one moves as far uphill or as little
downhill on $\hat{f}$ as possible. This procedure gives a path that follows
the highest gradient of $\hat{f}$ when approaching a mode, while it follows the
smallest gradient when it goes away from a mode (see e.g. Silverman 1986).
 The original method is due to Kittler (1976), while
 we are using a slightly modified version. We refer to the original Kittler
(1976) paper for the full set of theorems underlying this mapping. In this
section we only show the new version of the Kittler mapping we have introduced.
In fact the aim of the method was to
identify peaks in a general multivariate probability density field estimated
by the fixed kernel method (Parzen, 1962). As a consequence, the Kittler method
inherits the problems of the fixed kernels estimator of the probability density
(see discussion in paper I and references therein). At variance with the
original Kittler paper, we adopt
the density estimate $f_{ka}$
described in \S~2 and
the peak identification procedure described
in paper I and its extension to the general multivariate case outlined in
\S~2.
We are using the Kittler mapping as a useful tool for visualizing the
structure of a multivariate data set and to order a generally multimodal set of
points by non-increasing density within connected overdense regions.

 In practice, the Kittler
mapping (hereafter $KM$) is a particular rearrangement $k = KM(i)$
of the data sample
$D_{N} = \{ \vec{r}_{i}: i=1,\ldots,N  \}$:
\be
KM : D_{N} \rightarrow {\cal K}
\ee

 Let us suppose that the
probability density field associated with this set is estimated by
 eq.~(\ref{fk}).
 According to Kittler, it is possible to choose arbitrarily the first point
 of the sequence
${\cal K}=\{ \vec{r}_{k}: k=KM(i), i=1,\ldots,N \}$,
however we prefer to start with the highest probability
point:
\be
\vec{r}_{k=1}= \vec{r}_{i_{max}}
\ee
or $1 = KM(i_{max})$,
where $i_{max}$ is defined by:
\be
f_{ka}(\vec{r}_{i_{max}}) \geq f_{ka}(\vec{r}_{i}),
 \;\;\;\; \forall \; i \in [1,N].
\ee
 Then we can consider the set of points that are neighbours of
 $\vec{r}_{k=1}$.
Among these points we can choose the next point $\vec{r}_{k=2}$
of the sequence ${\cal K}$ as the
one having the highest value of the
density $f_{ka}(\vec{r})$, excluding the point corresponding to
 $\vec{r}_{k=1}$.
At the next step, all of the points which are neighbours of the last point
 $\vec{r}_{k=2}$
 of the sequence are also included as the set of neighbours among which
the search of the highest density point is made. The procedure is repeated in
this way for the other members of the set until no further neighbour is found.
 Since this may happen before all the data in the starting set has found
a place in the sequence ${\cal K}$,
we have introduced the convention that when the
procedure breaks, it restarts with the point not yet present within the
sequence that is the closest to the starting point
$\vec{r}_{k=1}$.
In order to completely define the procedure, we define two points
 $\vec{r}_{i}$ and
$\vec{r}_{j}$ to be neighbours if the distance between them is not
larger that the sum of the
sizes of their corresponding kernels $h_{i}+h_{j}$ appearing in eq.~(\ref{fk}):
\be
\mid \vec{r}_{i} - \vec{r}_{j} \mid  \leq  h_{i} + h_{j}
\label{dcrit}
\ee
similarly to the definition given by Kittler (1976). In order to keep the above
relation as simple as possible, we have considered the Kittler mapping only by
using the spherical kernel estimate of the density described by eq.~(\ref{fk}).
 It is
possible to extend the KM to non-isotropic kernels (eq.~[\ref{fw}]),
 at the cost of
longer computing time.

  By considering the peak to which $\vec{r}_{k=1}$ belongs,
 the above mapping
gives the sequence of points that would be obtained by cutting the
$f_{ka}(\vec{r})$
 hypersurface by hyperplanes of constant density at lower and
lower values. When another nearby peak is reached the sequence of points
reaches the highest density point following the steepest path and
 than re-descends
through the remaining points as for the previous peak.

  Once the sequence ${\cal K}$ is obtained,
it is possible to consider the plot:
$\{(k,f_{ka}(\vec{r}_{k})), k=KM(i), i=1,\ldots,N\}$ that
 shows in two dimensions
the structure in the multivariate data set, as outlined in full mathematical
detail by Kittler (1976).
Although this procedure may be not rewarding in two dimensions (see
Figs.~6 and 13) and is certainly not in one dimension, in our opinion it is
very
useful for visually examining data in three and higher dimensional spaces.
However it is worth stressing that the modality of the density  field
$f_{ka}(\vec{r})$ and in the
Kittler graph may not be exactly the same. This disagreement may occur
for some peak that has a member falling in the low probability tail and,
 in that
case, its distance from the nearest member of that peak may exceed the critical
value defined in eq.~(\ref{dcrit}).
 Because of this, we prefer to consider the
DEDICA method (\S 2) in order to define the systems present within the data
set.

 \section{Effects of structure and substructure}

 In this section we want to consider the effect that the presence of structure
 has on the estimate of a given parameter.

 Let us consider the following example. Suppose we have a set $S_{1}$
(see Fig.~9) of $N$
 points in a
 the $(x,y)$ space obtained by a random sampling of, say, a centrally symmetric
 and bivariate Gaussian with unit variance
centered on $(3,3)$
 and let $f_{1,ka}(x,y)$ be the density estimated according to
the method described in \S~2.1.
 We arrange the data set in a sequence ${\cal K}_{1}$
in order of non-increasing density:
\be
f_{1,ka}(x_{k},y_{k}) \geq f_{1,ka}(x_{k+1},y_{k+1})
\label{dsort}
\ee
 for $k=1,\ldots,N-1$. Let
us call $\sigma_{k,1}$ the value of the coordinate dispersion estimated from
the first $k$ points along the sequence ${\cal K}_{1}$.
 This is equivalent to saying
that $\sigma_{k,1}$ is the dispersion of the coordinates of the subset of
points
occupying the region defined by:
\be
 f_{1,ka}(x,y) \geq \delta_{k}
\label{overd}
\ee
where $\delta_{k} = f_{1,ka}(x_{k},y_{k})$. In Fig.~10a
 we show the plot of
$\sigma_{k,1}$ as a function of $k$ for
$S_{1}$.
 The application of DEDICA shows the presence of one structure without isolated
points.
 It can be seen that $\sigma_{k,1}$ grows quite smoothly with
$k$. The fluctuations are mainly due to noise in the density and $\sigma$
estimates.
 Let us consider now a second sample $S_{2}$ (see Fig.~9) containing
$N=100$ points
obtained by a random sampling of a centrally symmetric and
bivariate Gaussian with unit variance
centered on $(0,0)$. Let $f_{2,ka}(x,y)$ be the density estimate
obtained for the sample $S_{2}$.
In this case DEDICA indicates the presence of two significant peaks
(relative to a flat background)  and no
isolated points. However the two peaks are rather close to one another and
their
overlap is large. Both peaks have 50 members while there are 92
  members shared by the two peaks with a value of the parameter
${\cal L} = 11.5$.
 It is worth stressing the fact that the presence of two different peaks is
not due to a non-optimal shape of the kernels; in fact the kernels are
centrally symmetric so their shape is optimal in the sense of the Fukunaga
(1972) theorem (see \S 3). In this case the detected structure is due to
random fluctuations in $f_{2,ka}(x,y)$.
 If we sort the $S_{2}$ points in the same way as for $S_{1}$
(eq.[\ref{dsort}]), we obtain the sequence ${\cal K}_{2}$. The value of the
coordinate dispersion $\sigma_{k,2}$ along the ${\cal K}_{2}$ sequence grows
quite smoothly with $k$. Moreover it can be seen in Figs.~10a
and 11a that
$\sigma_{k,1}$ and $\sigma_{k,2}$ both follow a rather smooth dependence
 on $k$ although $\sigma_{k,2}$ is larger than $\sigma_{k,1}$. The
jack-knife estimates of the uncertainties\footnote{Hereafter
$\Sigma(p)$ indicates
the jack-knife estimate (Efron \& Tibshirani 1986)
of the dispersion of the quantity $p$.}
 $\Sigma(\sigma_{k,1})$ and $\Sigma(\sigma_{k,2})$ are similar
(see Fig.~10b and 11b).

 Finally, let us consider the sample $ S =  S_{1} \bigcup
  S_{2}$ (see Fig.~9).
If we arrange the data in order of non-increasing density as
in the previous example, the points  of the sequence jump continuously from
one system to the other (see Fig.~12a).  In order to avoid this problem and to
sort points by decreasing density without mixing points that belong to
overdense regions that are disconnected,
 we consider the following arrangement
(the Kittler sequence): the first $k=1$
point of the sequence ${\cal K}$
 is the one having the highest value of the density estimate, then from the
$k^{th}$ point the sequence moves along the $z = f_{ka}(x,y)$
 surface\footnote{Here $f_{ka}(x,y)$
is the estimate of the probability
density of the bimodal sample $S$.}
to the next point not yet listed in the sequence following the shallowest
 path when descending from a
density peak, while it follows the steepest path when rising to a peak.
In Fig.~12b
we show the sequence of the abscissae of the sample points along the Kittler
sequence. It can be seen that the sequence first spans the
overdense
 region defining the first peak then moves quickly to the top of the
 second peak and spans the remaining region of the second peak; finally it
covers the underdense region surrounding both peaks (see also Fig.~13).

 In Fig.~14a we show the plot of $\sigma_{k}$ versus $k$ along ${\cal K}$. It
can be seen that $\sigma_{k}$ grows smoothly until $k=60$ where the
sequence begins to include points belonging to the second peak. Then the value
of $\sigma_{k}$ and its dispersion $\Sigma(\sigma_{k})$
(Fig.~14b)
 rise quickly. The value of $\sigma_{k}$ finally
saturates after the overdense region of the second peak is covered.

 In conclusion it is possible to say that the presence of several well
separated systems within a sample may cause discontinuities in the estimate
of some structure-sensitive parameter along the Kittler sequence. On the other
hand small scale fluctuations of the probability density
 characterizing the
substructure within one system can cause the presence of different peaks
 close to one another and also some
minor fluctuations in the parameter
estimate.

\section{ An outline of the analysis}

 By exploiting the methods described in the previous sections we have
designed a procedure that analyses the structure of a data sample which
represents the three-dimensional positions of a set of point-like masses.

The analysis is divided in two parts. The first part
considers only the geometrical
structure ignoring the fact that the points we are considering form a dynamical
system. In the second part we analyse the structure and substructure by
considering their effects on the estimate of a given parameter along a sequence
of sample points ordered by non-increasing density within connected
overdense regions.

 Hence the final structure of the three-dimensional extension of DEDICA is the
following:
\begin{enumerate}

\item analysis  of the geometrical structure

  \begin{itemize}
     \item estimate the three-dimensional probability density function by the
           $f_{ka}(\vec{r})$
 defined in \S~2, and isolate the galaxy
           systems as peaks in $f_{ka}(\vec{r})$;
     \item estimate the cluster significance ${\cal S}_{\mu}$ of each system
         and the membership
            probability $P(i \in \mu)$ of each galaxy;
     \item apply the contact test;
 in the case that any systems are found to
  be in contact then apply a local whitening
in order to correct
        $f_{ka}(\vec{r})$ for locally non-isotropic kernels.
  \end{itemize}

\item  analysis of structure and substructure effects
   \begin{itemize}
       \item by using the density estimate $f_{ka}(\vec{r})$
get the Kittler
              sequence ${\cal K}$ and the plot of the KM:

 \be
\{ (r_{i},h_{i})\} \rightarrow \{ (k,f_{ka}(r_{k})) \}
\label{km}
\ee

        \item consider the following sequences: $M_(k)$, $\sigma_{V}(k)$
and $R_{V}(k)$ obtained by estimating the virial mass, velocity dispersion and
virial radius of the first $k$ galaxies within the Kittler sequence
${\cal K}$;
also consider the plot of $\sigma_{V}(k)$ versus $R_{V}(k)$ which should
indicate the presence of systems elongated along the line of sight (the
fingers-of-God effect);
use all these plots to infer information concerning the structure
of the sample.
\end{itemize}
\end{enumerate}
In the next section, we show the results obtained by applying the procedure
outlined here when applied to two galaxy clusters taken from the literature.

\section{A1656 (Coma)}

  This is a classical rich cluster. The data are taken from Kent \& Gunn
(1982) and were kindly provided to us in a computer readable form by
Girardi et al. (1994).
The list of the magnitude complete sample contains $N=337$ galaxies
brighter than 16.0. We have rejected the galaxies with unknown magnitude.
We want to
examine the clustering structure in this quasi-three dimensional space,
taking
$V$ as the radial polar coordinate.

In order to have homogeneous coordinates,
for the present cluster and for the following one,  we
transform the $(\alpha,\delta,V)$ set to the equivalent cartesian coordinates
$(x,y,z)$, by:

\be
\left\{ \begin{array}{l}
x = V \cos(\delta) \cos(\alpha) \\
y = V \cos(\delta) \sin(\alpha) \\
z = V \sin(\delta).
\end{array}
\right.
\ee

In Fig.~15 we plot the $(x,y)$ positions of the galaxies in this sample.
Contrary to the visual impression given by Fig.~15 and later by Fig.~24, we
have decided not to blindly whiten the data, but to apply the whitening
procedure only when the contact test is positive as is suggested by
the discussion in \S~4.
 By
applying the three-dimensional version of DEDICA to this data sample, we have
detected the presence of $\nu=56$ clusters plus $n_{0}=84$ isolated galaxies.
Among these clusters 22 have $0.90 \leq {\cal S}_{\mu} < 0.99$
 and are listed in table~\ref{ds_a1656}
 while 13 have ${\cal S}_{\mu}\geq 0.99$. The contact test shows the presence
of
only weak contact, moreover in no case have we obtained unimodal systems after
the merging. Because of this reason we do not consider the whitening of the
data.
Hence it seems that the A1656 sample contains a rich collection of
geometrically disjoint systems (see Fig.~16).
 This indication is supported also by the Kittler map (see Fig.~17)
where the peaks indicating the significant clusters are separated by
points at rather low density values. In fact from Figs.~16 and 17 it can
be seen that A1656 consists of a very dense system surrounded by several
satellite clusters at nearly $1/4$ or $1/5$ of the peak density (see
also Fig.~18).
  The first
two peaks in the Kittler map are due to the clusters $\mu=31$ and $\mu=32$
which
are also in  contact (see table~\ref{ct_a1656}).
These results suggest that A1656
has a  structure consisting of a bimodal core and an extended halo.

  A bimodal structure in the centre of the Coma cluster has also
been found in previous studies (see, for example,
the results obtained by
Fitchett \& Webster (1987), Mellier et al. (1988)).
 Our result supports the conclusions of Fitchett
\& Webster (1987) and is more general since it is not constrained to detect
only
bimodal structures
 as does the Lee statistics.
(see \S 1 and paper I).  The large number of local peaks found by
Mellier et al. 1988 is also in agreement with our results, although these
authors have used mainly bi-dimensional data.
 On the other hand, several authors (see e.g.
Dressler \& Schectman 1988; West \& Bothun 1990) have found no sign of
significant substructure in the Coma cluster. This disagreement is
probably due to the particular definition of the statistics which they have
adopted in order to detect substructures.

Concerning the comparison between the presence of structure
derived from optical data and the smooth appearance of the x-ray brightness
distribution, it is possible to note (Fitchett 1990) that x-ray emission from
the gas traces the gravitational potential in the standard hydrostatic
model. Since the galaxy  distribution is linked to the gravitational
potential through the Poisson equation, it is not unreasonable that the
x-ray gas distribution is smoother than the galaxy distribution.

 \begin{center}
{\bf Structure and substructure effects}
\end{center}

  In Figs.~19a,b we show the plot of the virial mass $\log M(k)$
and its dispersion $\Sigma[M(k)]$
 estimated following
 the Kittler sequence. It is possible to distinguish at least three regions in
$\log M(k)$.
 The first region goes from $k=1$ to $k \sim 34$. We call this
region the core of the cluster. It is mainly due to the presence of the two
groupings: $\mu=31$ that extends up to $k=18$, and $\mu=32$ that covers the
main
part of this region.
The smooth dependence of $M$ on $k$ passing from one
cluster to the other indicates that the presence of these
two clusters is probably due to substructure.
 We have also applied the pairwise Newton test (see e.g. Beers, Geller
\& Huchra 1982)
to these clusters. The
result was that
$$
\frac{V_{p}^{2} R_{p}}{0.76 G (M_{31} + M_{32})} \simeq 2.1
$$
 Hence this suggests that the two clusters are unbound.
 However this test assumes
that the clusters are point-like masses and this is probably not
reasonable
for
extended systems which are in contact such as $\mu=31$ and $32$.
The mass of the core is estimated as $\log M_{c} = 12.94 \pm 0.02$ in
solar units (here and in the following the error estimates are obtained by
using the jack-knife method).
 This value is roughly consistent with the upper limit of the
core mass estimated as $\log M_{c} \leq 12.7$ by Kent \& Gunn (1982). However,
we note that their estimate is model dependent and considers only the
galaxies within $3^{\circ}$ of the centre whereas we have considered
all of the galaxies and our mass estimate is model-independent.
Beyond $k \sim 34$ and up to $k \sim 300$ the slope of $M$ versus
$k$ changes significantly.
 The plot of
$\sigma_{v}(k)$ versus $R_{V}(k)$ (Fig.~20)
shows the presence of several
"fingers-of-God": regions where the value of the virial radius $R_{V}$ stays
roughly
constant, while the velocity dispersion $\sigma_{v}$ grows. The last finger
terminates roughly at $\sigma_{v} \sim 950 Km/s$ (i.e. at $ k \sim 300$).
This suggests to us that probably the outer limit of the cluster occurs at this
value of $k$.

These results suggest that this region may be composed of
satellite clusters distributed around the
dense central core of A1656. Let us call this region the halo. The mass of the
core and halo is estimated as $\log M_{c+h} = 14.904 \pm 0.003$
 in solar units. The
best-fit estimate of the mass of the cluster given by Kent \& Gunn (1982) is
$\log M_{c+h} = 15.10$, again not very different from our estimate.
Moreover, by using both optical and x-ray data, Hughes (1989) obtained
a value of $M = ( 1.85 \pm 0.24) \times 10^{15} h_{50}^{-1} M_{\odot}$
corresponding to $\log M = 14.97$ in our units. This is again in rather good
agreement with our estimate.

We must state clearly that the values of the virial mass which we quote
are from blind application of the virial theorem.
The mass estimates are
more a rough number computed in order to compare our results with those
in the previous literature.  We plan to explore the possibility of
exploiting the structure to infer a self-consistent
dynamical model of these systems without  being constrained by the
assumptions of the virial theorem.  A similar piece of work  for
spherically symmetric systems is that of Merritt \& Shaha (1993).

Finally, for
$k \geq 300$, the slope of $M(k)$  increases and this may be due to
background or foreground galaxies which are not linked to the
cluster.

 The above discussion is confirmed by the plots of the velocity dispersion
$\sigma_{V}(k)$ and of the virial radius $R_{V}(k)$ versus $k$ (Figs.~21a
and 22a)
 as well as
by the plots of $\Sigma[\sigma_{V}(k)]$ and $\Sigma[R_{V}(k)]$ versus $k$
(Figs.~21b and 22b) along the Kittler
sequence. In fact from $\sigma_{V}(k)$ (Fig.~21a) it is possible to notice the
presence of a region with low dispersion and relatively low value of the
jack-knife uncertainty $\Sigma(\sigma_{V})$. This region extends out to
$k = 34$ where $\sigma_{V}(k)$ versus $k$ shows a quick decrease in slope and a
quick rise of $\Sigma(\sigma_{V})$.  Within this region the virial radius
grows until $k \sim 20$, beyond this value $R_{V}$ levels to $\sim 1 h \; Mpc$.
This is the central $( k \leq 34)$
high density, low dispersion region we can call the core.

 For $k\geq 34$ the values of both $\sigma_{V}$ and $R_{V}$ grow quite smoothly
 with $k$ indicating that as we move along the Kittler sequence towards
galaxies
falling in regions of lower and lower density, the system expands both in the
radian direction and in the plane normal to the line of sight. It turns out
that
along the line of sight the structure of the sample is unimodal and
characterized by a single peak at $7000 \; Km/s$ (see Fig.~23). As $k$ grows
along ${\cal K}$ we move from the high density peak of $f(V)$ towards the low
density wings of $f(V)$. This fact causes the value of $\sigma_{V}$ to grow
smoothly with $k$ and $\Sigma(\sigma_{V})$ to decrease. On the other hand, the
structure on the plane tangential to the celestial sphere is multimodal and
this
causes the value of $\Sigma(R_{V})$ to fluctuate quite widely. This occurs
until roughly $k \sim 150-170$. Beyond this value of $k$ the value of $R_{V}$
stops growing and $\Sigma(\sigma_{V})$ has a minimum. This can be interpreted
by
saying that for larger $k$ (and hence lower density threshold) the system
expands mainly along the line of sight and hence the projected size of the
system estimated by $R_{V}$ stays constant while its uncertainty
$\Sigma(R_{V})$ decreases.
 On the other hand both $\sigma_{V}$ and $\Sigma(\sigma_{V}
)$ grow smoothly
due to the fact that in this part of the Kittler sequence the galaxies
 fall in the low $f(V)$ wings. Finally at roughly $k \sim 300$ $R_{V}$ begins
to grow with $k$, moreover both $\sigma_{V}$ and $\Sigma(\sigma_{V})$ grow
rather steeply with $k$. This is a reasonable indication that
we are now out of the halo of the
Coma Cluster and including field galaxies.

\section{The Cancer cluster}

  As an example of a multimodal cluster, we have considered the Cancer cluster.
The data  are taken from Bothun et al. (1983) where
 positions and redshifts are listed for $N=123$ galaxies whose positions are
shown in Fig.~24.
 The three-dimensional geometrical
analysis of DEDICA shows the presence of $\nu=24$ clusters and
$n_{0}=32$ isolated galaxies (see Fig.~25).
 Among these clusters there are 6 with
 $ 0.80 \leq {\cal S}_{\mu} < 0.90$, 6 with $0.90 \leq{\cal S}_{
\mu} < 0.99$ and 5 with ${\cal S}_{\mu}\geq 0.99$ (see table~\ref{ds_cancer}).
 We must
stress the fact that the galaxies classified by DEDICA as clusters $\mu=10$ and
$\mu=16$ are not considered in the clustering analysis by Bothun et al. (1983)
since they clearly appear to be foreground projections. However we decided to
include these galaxies in the sample and let DEDICA show that they are actually
foreground objects. The contact test shows the presence of weak
asymmetric contact of strength ${\cal L}=0.12$ between $\mu=10$ and $\mu=16$.
Hence the clusters found by DEDICA are rather well disjoint.
As for A1656 the merging and whitening of the overlapping systems do not give
unimodal systems. Hence we do not consider whitened data.
The multimodal structure of this sample is well shown by the Kittler map in
Fig.~26 where the highest peaks are associated with the significant clusters
and
appear to be separated by low density regions (see also Fig.~27).

Comparing the results obtained by DEDICA and those of Bothun
et al. (1983) it is possible to say that there is only qualitative agreement.
In fact Bothun et al. (1983) found a large group which they call $A$ with
$n_{A}= 38$ members plus four more groups with $n_{B}=6$, $n_{C}=9$, $n_{D}=7$
and $n_{E}=10$ members.
With the exception of $\mu=10$, all the significant groups (i.e. those having
${\cal S}_{\mu} \geq 0.99$) found by DEDICA (namely $\mu=11, 12, 13, 19$)
are contained within the $A$ group of Bothun et al. (1983). Hence the $A$ group
shows the presence of significant substructure according to our
analysis. A similar conclusion holds for the remaining groups $B$, $C$,
$D$ and $E$. In fact  they are superpositions of one or two marginally
significant  groups (i.e. having $0.90 \leq {\cal S}_{\mu} <0.99$) and of
some nearby fluctuations of the density field.

 \begin{center}
{\bf  Structure and substructure effects}
 \end{center}

  We have performed the
Kittler mapping for several clusters assuming the highest density galaxy of
each
cluster as the starting point of the Kittler sequence.

  The easiest cluster to begin with is the foreground cluster $\mu=10$. The
Kittler plot $M(k)$ shows a very sharp break in slope at $k=10$,
 while there is
a smooth change of $M(k)$ passing from $k=6$ that marks the
 end of $\mu=10$ to
the range $k\in[6,10]$ that is occupied by $\mu=16$. This suggests that
 the existence of two different structures $(\mu=10$ and $\mu=16)$ instead of
only one is due to substructure.
 The pairwise Newton test between these clusters indicates that
they are mildly unbound. However due to the contact between these structures
the
hypotheses underlying this test are probably not satisfied.
A blind application of the virial theorem gives the mass of this unit
($C_{10} \bigcup C_{16}$)
 as $\log M_{10+16}/M_{\odot} = 12.10 \pm 0.07$.

 The remaining four significant structures $\mu= 11$, $12$, $13$, and $19$ are
well disjoint and  satisfy the pairwise Newton test. Moreover in the
$M(k)$ plot (Fig.~28a) the curve seems smooth
out to $k = 22$ while beyond this
limit the slope of $M(k)$ rises.
 A blind application of the virial estimator would give a
mass of $\log(M/M_{\odot})=13.51 \pm 0.02$.

The plot of $\sigma_{V}(k)$ versus $R_{V}(k)$ is shown in Fig.~29. It is
possible to notice the lack of prominent vertical sequences of points (the
finger-of-God effect) supporting the multimodal structure of this sample.

  The group $B$ is found by DEDICA to be composed of the weakly
significant cluster $\mu=14$ together with ${\cal S}_{14} = 0.86$ and
three isolated galaxies.
According to the Kittler mapping, the
$M(k)$ curve shows a break at $k=6$ with $\log M(k=6) = 12.95
\pm 0.09$. The six galaxies are the members of $\mu=14$ plus three
isolated galaxies, however these additional members are not the same
galaxies  that Bothun et al. (1983) attribute to this structure.

  The group $C$ appears to comprise galaxies attributed by DEDICA to $\mu=17$,
$18$ and $20$. The most significant cluster among these is $\mu=18$ that
has $\log M(k=5) = 11.7 \pm 0.3$.

 The group $D$ is made essentially by the clusters $\mu=4$ and $\mu= 8$. These
two clusters are indicated by our analysis
as probably being subunits of a larger structure
with $\log M(k=8) = 12.8 \pm 0.1$.

  Finally the group $E$ corresponds in part to the clusters $\mu=7$ and
$\mu=9$ which
seem to form a bound structure  with $\log M(k=12) =
12.80 \pm 0.07$.

 We note that the values of the mass that we obtain for the
clusters mentioned
are almost two orders of magnitude smaller than the estimates given by
Bothun et al. (1983). This discrepancy is due mainly to the presence of
interlopers in the Bothun et al. (1983) groups and slightly to a difference
 in the definition of virial radius (see the definition in e.g. Pisani et al.
1992).

  Concerning the plots of $\sigma_{V}$, $R_{V}$ and $\log(M)$ with their
uncertainties $\Sigma(\sigma_{V})$, $\Sigma(R_{V})$ and $\Sigma(\log M)$ versus
$k$ we can say that the virial mass shows the same shape as in the case of
A1656, but with a very different slope. In
fact at $k=120$ the mass of Cancer is two orders of magnitude larger than in
the A1656 sample. Moreover, there is some indication of fluctuation in the
slope at $k=15$ and $22$. The main differences relative to the A1656 sample
are linked to the velocity structure and to  $\sigma_{V}(k)$ and $R_{V}(k)$.
 In Fig.~30 we show the velocity structure of Cancer. It is possible to notice
the presence of $5$ peaks. Two of these are highly significant (${\cal S} \geq
0.999$); the first is centered at $2009 \; Km/s$ and contains $11$ members
while
 the second is centered at $4887\; Km/s$ and contains $91$ members. Moreover
in the first peak of $f(V)$ we find the clusters classified as $\mu=10$ and
$\mu=16$ by the three-dimensional analysis, while in the peak centered at
$4887 \; Km/s$ we find $\mu=11$, $12$, $13$ and $19$. The velocity dispersion
$\sigma_{V}$ grows quite smoothly with $k$ (Fig.~31a).
 The value of $\Sigma(\sigma_{V})$
has a minimum at $k \sim 50$ (Fig.~31b).
 For $k \geq 90$ the Kittler sequence includes
members of different peaks causing both $\sigma_{V}$ and $\Sigma(\sigma_{V})$
to
increase. Moreover $R_{V}$ is rather constant for $k\leq 20$, while it grows
almost steadily for the remaining part of the sequence (Fig.~32a).
 A short plateau may be
noticed centered at $k\sim 100$. Hence apart from the first 20 members of
${\cal K}$ for which the system develops along the line of sight, the Kittler
sequence shows that the structure develops in both directions parallel and
normal to the line of sight. Moreover there is evidence of significant
structure
both in velocity and in plane normal to the line of sight as also
suggested by the fluctuations in $\Sigma(R_{V})$ (see Fig.~32b).

 In conclusion DEDICA shows the presence of at least two
independent structures. None of these are free of significant
 substructure.

  We may say that our analysis is in qualitative agreement with
the results obtained by Bothun et al (1983) since we detect the presence of
several geometrically and probably also dynamically unrelated structures.
However  the details of the clustering pattern that we found is
rather different from that found by Bothun et al. (1983).

\section{Summary and conclusions}

  In this paper we have confronted the problem of extending to the
general multivariate case the clustering algorithm called DEDICA which
was introduced by Pisani (1994) for univariate problems.
The extension is in principle straightforward as several
numerical simulations have shown. Moreover we have proposed a method that
allows us to estimate the locally optimal metric and which hence improves the
performance of the density estimator by reducing the noise in the
gradient of the density.
The method proposed basically
rests on the assumption that a highly elongated
structure is broken into several subsystems due to the sphericity of the
kernels. These subsystems lie quite close to one another.
We have introduced
a non-parametric  estimate of the contact between these subsystems.
After the merging of the significant systems that are in mutual contact we
can estimate the locally optimal metric and significantly reduce the noise.
 Several tests on Monte Carlo simulations have supported
the effectiveness of this method for both two and three-dimensional data
samples.

  Parallel to this noise reduction method, we have used a slightly
modified version of a procedure called Kittler mapping for building a
sequence of the data in the sample allowing us to show in a two
dimensional plot the structure of a multidimensional data sample. We
have used the Kittler mapping both to support the geometrical analysis
of DEDICA and to obtain information about the effects of the structure
 detected within the
sample. In fact, we have shown by using some examples that along the
Kittler sequence the estimate of a given parameter (such as, for
example, the coordinate dispersion $\sigma$)
  as a function of the number of particles
considered, makes a quick change in slope when significant and large-scale
 structures is present. On the contrary the chosen parameter changes quite
smoothly if significant but small-scale structure is present within the sample.
We can say that a sample has substructure if only small-scale structure is
present, while it has structure if large-scale structure is  present.
Further application of the Kittler mapping
for dynamical purposes on larger sample
is in progress.

  In summary we have presented a general method of cluster analysis
which is based on the estimate of the probability density of the data
sample and which satisfies the following requirements:

\begin{itemize}
\item it does not require assumptions to be made concerning the number
of members or any
other feature of the systems it is designed to look for;
\item it gives an estimate of the statistical significance of each system it
detects:  this is a useful number in order to distinguish likely systems
from noise fluctuations;
\item it gives an estimate of the membership probability for each point in the
data sample to each detected system; this quantity can indicate the presence of
interlopers within the detected systems;
\item
finally we have suggested a possible way
 to analyse the effects of structure and substructure
on the estimate of the cluster's  parameters;

\end{itemize}

  In this framework we think that the disagreement concerning the
definition, presence and relevance of structure and substructure within
galaxy clusters (compare e.g. Fitchett \& Webster 1987 with West \&
Bothun 1990) can be faced successfully.

In order to illustrate the performance of the method,
we have applied the three-dimensional version of DEDICA to two different
clusters: a rich unimodal cluster (Abell 1656, namely Coma) and a
multimodal cluster (Cancer). Our results indicate Coma as a
 core-halo structure formed by a rather large number
of geometrically distinct structures.
 For Coma, a rough quantitative agreement is obtained with
previous studies based on optical
(e.g. Kent \& Gunn 1982) and both optical and x-ray (Hughes 1989) data.
 On the other
hand, the Cancer cluster results
 composed of at least two
distinct structures both with substructure. In
the case of Cancer, only qualitative agreement is obtained with the
previous literature (Bothun et al. 1983).  We emphasize that our
estimate of the clustering structure is obtained without making any
assumption about the structures or any model concerning their
properties.

 Currently ongoing and future observational projects will provide   new and
rich data samples of galaxy clusters particularly suitable for the analysis
presented here. Moreover we plan to use the results of the geometrical
structures  to infer a self-consistent dynamical model of these systems in
a very general way without being necessarily constrained by the assumptions
of the virial theorem.

 We plan to extend the application of DEDICA to large scale galaxy
samples in order to analyse structures on larger scales than galaxy clusters.
Moreover, as an application of the non-parametric and hence
model-independent estimate of the density we plan to estimate the topological
properties of constant density surfaces estimated from rich galaxy samples.

\newpage
\begin{center}
{\large ACKNOWLEDGMENTS}
\end{center}

 This work is DEDICA-ted to my wife Antonella. I wish to thank for very
fruitful
discussion and friendly support  A. Biviano, G. Evrard,
G. Fasano,
 M. Girardi, G. Giuricin,
F. Lucchin,
F. Mardirossian, S. Matarrese, Y. Mellier, M. Mezzetti, J. Miller,
  L. Moscardini,  P.L. Monaco, F. Murtagh,
 M. Plionis, R. Stark, R. Vio.

 This work has been supported by the Ministero Italiano della Ricerca
Scientifica e Tecnologica.



\newpage

\begin{center}

{\large \bf List of figures}

\end{center}

\begin{description}
\item[Fig. 1]  Monte Carlo simulations of a bivariate density field. The
solid line contours indicate the isoplethes (namely level curves of constant
probability density)
of the spherical model field
$F(x,y)$ while
the dashed line contours indicate the kernel estimate $f_{ka}(x,y)$
(eq.~[\ref{fk}]) obtained by using kernels with the optimal metric. The number
of simulations is $N_{s} = 1000$ each with $N_{p} = 500$ points. The two
central crosses indicate the positions of the peaks of $F(x,y)$
 and $f_{ka}(x,y)$.
 The contours are constant levels of density corresponding
to 0.75, 0.50, 0.25 and 0.10 times the peak value of the model field $F(x,y)$.

\item[Fig. 2]  Monte Carlo simulations of a bivariate density field. The
solid line contours indicate the isoplethes of the elliptical model field
$F(x,y)$ with axis ratio $a/b=0.1$ and position angle $\vartheta = 45^{\circ}$,
while
the dashed line contours indicate the kernel estimate $f_{ka}(x,y)$
(eq.~[\ref{fk}]) obtained by using kernels with a spherical shape and hence a
 non-optimal metric. The number
of simulations is $N_{s} = 1000$ each with $N_{p} = 500$ points. The two
central crosses indicate the positions of the peaks of $F(x,y)$ and
 $f_{ka}(x,y)$.
 The contours are constant levels of density corresponding
to 0.75, 0.50, 0.25 and 0.10 times the peak value of the model field $F(x,y)$.
It can be seen that the isoplethes of $f_{ka}(x,y)$ are
slightly rounder that the isoplethes of $F(x,y)$.

\item[Fig. 3] An example of Monte Carlo simulation (MC)
of a bivariate elliptical
 density  field with axis ratio  $a/b=0.1$ and position angle $\vartheta =
45^{\circ}$ with $100$ points superposed onto a flat field with $100$ points.

\item[Fig. 4] The clusters identified by DEDICA. Each point is labelled by the
sequential number of the cluster to which
 it belongs (see column 1 in Tab.~2).

\item[Fig. 5] The adaptive kernel estimate $f_{ka}(x,y)$ obtained from the
MC data shown in Fig.~3.

\item[Fig. 6] The Kittler mapping (see \S~5) $(k,f_{ka}(x_{k},y_{k}))$
obtained from the density estimate
of the $MC$
data shown in Fig.~3. The presence of several close peaks
within the first $k \sim 100$ points apparently superposed on a larger scale
structure is consistent with the results of the contact test.

\item[Fig. 7] The values of the membership probability $P( i \notin 0) = 1 -
P(i \in 0)$ for the sample shown in Fig.~3. Here the probability density is
estimated by a centrally symmetric gaussian kernel. The contours refer to
constant probability levels at 0.90 (outer dotted line), 0.95 (middle
solid line) and 0.99 (inner dashed line).

\item[Fig. 8] The adaptive kernel estimate $f_{kaw}(x,y)$ corrected for locally
anisotropic kernels (eq.~[\ref{fw}])
obtained from the MC
data shown in Fig.~3.

\item[Fig. 9] Positions of the $S=S_{1} \bigcup S_{2}$ sample (see \S~6)
made by the
superposition of two Monte Carlo simulations of a bivariate Gaussian with
central symmetry, unit variance and centered on (3,3) for $S_{1}$ (crosses),
 and on
(0,0) for $S_{2}$ (open squares).

\item[Fig. 10] a) The coordinate dispersion $\sigma_{k,1}$ versus $k$ along
the Kittler sequence ${\cal K}_{1}$ for the $S_{1}$ sample. b) The jack-knife
estimate of the uncertainty in the coordinate dispersion $\Sigma(\sigma_{k,1})$
versus $k$ along ${\cal K}_{1}$.

\item[Fig. 11] a) The coordinate dispersion $\sigma_{k,2}$ versus $k$ along
the Kittler sequence ${\cal K}_{2}$ for the $S_{2}$ sample. b) The jack-knife
estimate of the uncertainty in the coordinate dispersion $\Sigma(\sigma_{k,2})$
versus $k$ along ${\cal K}_{2}$.

\item[Fig. 12] The sequence of abscissae of the $S$ data points
obtained by a) ordering the data by non-increasing density and b) following the
Kittler sequence.

\item[Fig. 13] The Kittler mapping $(k,f_{ka}(x_{k},y_{k}))$
obtained from the density estimate
of the $S$ data shown in Fig.~9
 The presence of a double peak
within the second
structure is consistent with the results of the contact test.

\item[Fig. 14] a) The coordinate dispersion $\sigma_{k}$ versus $k$ along
the Kittler sequence ${\cal K}$ for the $S= S_{1} \bigcup S_{2}$
 sample. b) The jack-knife
estimate of the uncertainty in the coordinate dispersion $\Sigma(\sigma_{k})$
versus $k$ along ${\cal K}$.

%
\item[Fig. 15] The A1656 data sample:
 projection of the galaxy
positions onto the $(x,y)$ plane.

\item[Fig. 16] The A1656 data sample:
 projection onto the $(x,y)$ plane of the
density estimate $\int f_{ka}(x,y,z) dz$.

\item[Fig. 17] The A1656 data sample:
 Kittler
mapping $(k,f_{ka}(x_{k},y_{k},z_{k}))$ for the full data sample.

\item[Fig. 18] The A1656 data sample:
 isoplethe contours of
 $\int f_{ka}
(x,y,z)dz$ for the main central part of the data distribution
namely:
$x (Km/s) \in [-8000,-3000]$, $y (Km/s) \in [-2400,-400]$. The contours
show the constant density levels corresponding to the following fractions of
the peak: 0.1, 0.25, 0.5, 0.75.

\item[Fig. 19] Plot of values estimated along
along the Kittler sequence of the A1656 data sample for a)
the decimal logarithm of the virial mass $\log M(k)$, and b) the
 jack-knife estimate of the dispersion $\Sigma[\log M(k)]$.

\item[Fig. 20] Plot of the velocity dispersion $\sigma_{V}(k)$ versus the
virial radius $R_{V}(k)$ along the Kittler sequence of the A1656 data sample.
The presence of several "fingers-of-God" can be seen.
 The crosses indicate the positions of the pairs
$(\sigma_{V}(k),R_{V}(k))$,
while the vertical and horizontal bars indicate the jack-knife estimate of the
three sigma error band.

\item[Fig. 21] Plot of values estimated along
along the Kittler sequence of the A1656 data sample for a)
the velocity dispersion $\sigma_{V}(k)$, and b) the
 jack-knife estimate of the dispersion $\Sigma[\sigma_{V}(k)]$.

\item[Fig. 22] Plot of values estimated along
along the Kittler sequence of the A1656 data sample for a)
 the virial radius $R_{V}(k)$, and b) the
 jack-knife estimate of the dispersion $\Sigma[R_{V}(k)]$.

\item[Fig. 23] Estimated density of radial velocity $f(V)$ for the A1656
sample.

\item[Fig. 24] The Cancer data sample:
 projection of the galaxy
positions onto the $(x,y)$ plane.

\item[Fig. 25] The Cancer data sample:
 projection onto the $(x,y)$ plane of the
density estimate $\int f_{ka}(x,y,z) dz$.

\item[Fig. 26] The Cancer data sample:
  Kittler
mapping $(k,f_{ka}(x_{k},y_{k},z_{k}))$ for the full data sample.

\item[Fig. 27] The Cancer data sample:
 isoplethe contours of
 $\int f_{ka}
(x,y,z)dz$ for the main central part of the data distribution,
namely:
$x (Km/s) \in [-5000,-1500]$, $y (Km/s) \in [2000,7000]$. The contours
show the constant density levels corresponding to the following fractions of
the peak: 0.1, 0.25, 0.5, 0.75.

\item[Fig. 28] Plot of values estimated along
along the Kittler sequence of the Cancer data sample for a)
the decimal logarithm of the virial mass $\log M(k)$, and b) the
 jack-knife estimate of the dispersion $\Sigma[\log M(k)]$.

\item[Fig. 29] Plot of the velocity dispersion $\sigma_{V}(k)$ versus the
virial radius $R_{V}(k)$ along the Kittler sequence of the Cancer data sample.
No prominent "finger-of-God" is evident.
 The crosses indicate the positions of the pairs
$(\sigma_{V}(k),R_{V}(k))$,
while the vertical and horizontal bars indicate the jack-knife estimate of the
three sigma error band.

\item[Fig. 30] Estimated density of radial velocity $f(V)$ for
 the Cancer sample.

\item[Fig. 31] Plot of values estimated along
along the Kittler sequence of the Cancer data sample for a)
the velocity dispersion $\sigma_{V}(k)$, and b) the
 jack-knife estimate of the dispersion $\Sigma[\sigma_{V}(k)]$.

\item[Fig. 32] Plot of values estimated along
along the Kittler sequence of the Cancer data sample for a)
 the virial radius $R_{V}(k)$, and b) the
 jack-knife estimate of the dispersion $\Sigma[R_{V}(k)]$.

\end{description}


\begin{table}
\caption
[]
{Values of $\Delta$ (eq.~[\ref{delta}]) that
result from the $N_{s}=1000$
 Monte Carlo simulations test with $N_{p}=500$ of the
performance of the density estimators for three values of the axis ratio
 $a/b$.}
\label{tab_delta}
\begin{center}
\begin{tabular}{cccc}
 &  $a/b=1$ & $a/b=0.5$ & $a/b=0.1$ \\
 &  &  &   \\
min\{$\Delta$\} & $1.42 \times 10^{-5}$ & $4.34 \times 10^{-5}$ &
 $8.26\times 10^{-4}$ \\
max\{$\Delta$\} & $5.87\times 10^{-4}$ & $9.37 \times 10^{-4}$ &
$5.64\times 10^{-3}$  \\
median\{$\Delta$\} & $5.74\times 10^{-5}$ &  $1.00 \times 10^{-4}$ &
$1.43\times 10^{-3}$ \\
$\Delta_{0.25}$ & $2.91\times 10^{-5}$& $7.88 \times 10^{-5}$ &
$1.26\times 10^{-3}$  \\
$\Delta_{0.75}$ & $4.93\times 10^{-5}$ & $1.43 \times 10^{-4}$ &
$1.65\times 10^{-3}$ \\
mean\{$\Delta$\}& $4.81\times 10^{-5}$ & $1.30 \times 10^{-4}$ &
$1.50\times 10^{-3}$ \\
st.dev.\{$\Delta$\} & $6.28\times 10^{-5}$ & $1.52 \times 10^{-4}$ &
 $1.58\times 10^{-3}$ \\
\end{tabular}
\end{center}
\end{table}

\begin{table}[h]
\caption
[]
{The cluster parameters and significance for the $MC$ simulated data
 shown in Fig.~3 (see \S~4).}
\label{mc_ds_1}
\begin{center}
\begin{tabular}{rrrrrrr}
$\mu$ &  $n_{\mu}$ & $\bar{x}$ & $\bar{y}$ &
  $\sigma_{x}$ & $\sigma_{y}$ & ${\cal S}_{\mu}$ \\
  &   &   &   &   &   &  \\
0 & 0 & - & - & - & - & - \\
1 & 14 & 0.78 & 0.77 & 0.11& 0.09& $>0.999$ \\
2 & 10 & -0.68 & -0.56 & 0.08 & 0.04 & $>0.999$ \\
3 & 2 & -3.06 & 1.40 & 0.06& 0.15& 0.466 \\
4 & 2 & 1.31 & -0.98 & 0.07 & 0.14& 0.492 \\
5 & 3 & 1.61 & 1.54 & 0.02 & 0.01 & 0.981 \\
6 & 3 & -0.44 & 1.52 & 0.23 & 0.01& 0.856 \\
7 & 6 & -1.67 & 1.13 & 0.19 & 0.27 & 0.716 \\
8 & 9 & 1.19 & 1.14 & 0.12 & 0.12 & $>0.999$ \\
9 & 26 & -0.05 & -0.04 & 0.13 &0.16 & 1.000 \\
10 & 4 & 1.06 & -2.03 & 0.15 & 0.33 & 0.651 \\
11 & 3 & 0.23 & -2.81 & 0.17 & 0.17 & 0.755 \\
12 & 14 &  -0.99& -0.89 & 0.21 & 0.12& $>0.999$ \\
13 & 4 & -0.59 & -2.42 & 0.11 & 0.26& 0.811 \\
14 & 14 & 0.44 & 0.40 & 0.11 & 0.12 & $>0.999$ \\
15 & 6 & -1.02 & 0.30 & 0.22 & 0.29& 0.857 \\
16 & 6 & -2.08 & 0.40 & 0.25 & 0.24& 0.747 \\
17 & 19 & -0.42 & -0.38 & 0.08 & 0.14 & 1.000 \\
18 & 3 & -2.34 & -0.86 & 0.12 & 0.14& 0.605 \\
19 & 12 & 0.22 & -0.69 & 0.35 & 0.27 & 0.991 \\
20 & 3 & 1.49 & 0.36 & 0.10 & 0.20 & 0.745 \\
21 & 3 & 0.23 & 1.15 & 0.06 & 0.28 & 0.682 \\
22 & 5 & -0.10 & -1.61 & 0.31 & 0.07 & 0.819 \\
23 & 9 &  -2.69 & -2.04 & 0.31 & 0.50 & 0.734 \\
24 & 3 &  -2.89 & 0.63 & 0.14 & 0.19 & 0.792 \\
25 & 2 &  -2.88 & -0.34 & 0.14 & 0.15 & 0.782 \\
26 & 13 & -1.71 & -2.08 & 0.21 & 0.50& 0.956 \\
\end{tabular}
\end{center}
\end{table}


\begin{table}[h]
\caption
[]
{The contact among clusters of the Monte Carlo simulation shown in Fig.~3
(see \S 4).}
\label{ct_1}
\begin{center}
\begin{tabular}{rrr}
$\mu,\lambda$ & n & ${\cal L}(\mu,\lambda)$\\
                      &   &    \\
$1 \leftrightarrow 8$ & 7 & 0.81\\
$1 \leftrightarrow 14$ & 10 & 1.52\\
$2 \leftrightarrow 17$ & 15 & 2.45\\
$2 \leftrightarrow 12$ & 8& 0.65 \\
$9 \leftrightarrow 17$ & 16& 1.52\\
$9 \leftrightarrow 14$ & 10& 1.01\\
$9 \rightarrow 19$ & 1 & 0.09 \\
$17 \rightarrow 19$ & 1 & 0.04 \\
\end{tabular}
\end{center}
\end{table}



\begin{table}[h]
\caption
[]
{The mean parameters and significance of the clusters in the
 $A 1656$ region. Only the clusters with ${\cal S}_{\mu} \geq 0.95$
are reported}
\label{ds_a1656}
\begin{center}
\begin{tabular}{rrrrrr}
 $\mu $ & $n_{\mu}$ & $\bar{V}$ & $\sigma_{V}$ & $\log(M) $ & ${\cal
S}_{\mu}$\\
  &  &  &  &  &   \\
 0 & 84 & - & -  & - & -  \\
 4 & 3 & 6637. & 56.15 & 11.31 & 0.965  \\
 11 & 5 & 1186. & 39.30 & 11.54 & 0.999  \\
 15 & 3 & 6872. & 35.10 & 11.84 & 0.964  \\
 19 & 4 & 7543. & 75.12 & 12.06 & 0.966  \\
 23 & 4 & 6175. & 36.34 & 11.91 & 0.966  \\
 25 & 5 & 7228. & 21.57 & 11.26 & 0.998  \\
 27 & 3 & 5516. & 10.01 & 10.68 & 0.969  \\
 28 & 6 & 8009. & 25.11 & 11.47 & 0.998  \\
 29 & 3 & 7703. & 16.64 & 11.20 & 0.958  \\
 31 & 18 & 6837. & 52.78 & 11.75 & 1.00  \\
 32 & 8 & 6671. & 39.65 & 11.73 & $> 0.999$  \\
 34 & 5 & 5369. & 52.82 & 11.93 & 0.982  \\
 36 & 3 & 7194. & 14.47 & 10.67 & 0.994  \\
 37 & 4 & 7604. & 28.71 & 11.49 & 0.990  \\
 39 & 18 & 7556. & 51.55 & 12.16 & $>0.999$  \\
 40 & 5 & 8179. & 47.91 & 12.15 & 0.962  \\
 42 & 6 & 6026. & 47.38 & 11.90 & 0.998  \\
 43 & 7 & 5663. & 52.39 & 12.13 & 0.999  \\
 44 & 17 & 7068. & 70.29 & 12.49 & $> 0.999$  \\
 46 & 7 & 6375. & 66.07 & 12.12 & 0.995  \\
 47 & 13 & 7834. & 68.77 & 12.43 & $> 0.999$  \\
 48 & 3 & 6642. & 12.12 & 11.00 & 0.955  \\
 49 & 9 & 5888. & 56.43 & 12.25 & 0.998  \\
\end{tabular}
\end{center}
\end{table}

\begin{table}[h]
\caption
[]
{The contact among clusters in the A1656 region.}
\label{ct_a1656}
\begin{center}
\begin{tabular}{rrr}
$\mu,\lambda$ & n & ${\cal L}(\mu,\lambda)$ \\
                      &   &     \\
$29 \leftrightarrow 37$ & 2 & 0.28 \\
$25 \leftrightarrow 36$ & 2 & 0.20 \\
$31 \leftrightarrow 32$ & 3& 0.20 \\
$18 \rightarrow 39$ & 1 & 0.15  \\
$29 \rightarrow 47$ & 1& 0.01 \\
\end{tabular}
\end{center}
\end{table}



\begin{table}[h]
\caption
[]
{The mean parameters and significance of the clusters in the Cancer region}
\label{ds_cancer}
\begin{center}
\begin{tabular}{rrrrrr}
 $\mu $ & $n_{\mu}$ & $\bar{V}$ & $\sigma_{V}$ & $\log(M) $ & ${\cal
S}_{\mu}$\\
  &   &      &      &      &   \\
 0 & 32  &  -    &  -    &  -    &  - \\
 1 & 3 & 4147. & 57.42 & 11.60 & 0.947 \\
 2 & 2 & 7501. & 41.01 & 12.24 & 0.643 \\
 3 & 2 & 4629. & 38.89 & 12.27 & 0.558 \\
 4 & 5 & 3512. & 97.02 & 12.56 & 0.948 \\
 5 & 3 & 4281. & 35.23 & 12.03 & 0.813 \\
 6 & 2 & 5706. & 36.77 & 11.85 & 0.834 \\
 7 & 6 & 4153. & 81.72 & 12.38 & 0.979 \\
 8 & 3 & 3763. & 43.88 & 11.80 & 0.791 \\
 9 & 3 & 3911. & 57.50 & 12.31 & 0.736 \\
 10  & 6 & 2130. & 39.50 & 11.53 & 0.999 \\
 11 & 9 & 5042. & 107.37 & 12.43 & $> 0.999$ \\
 12 & 5 & 4837. & 18.27 & 10.91 & $> 0.999$ \\
 13 & 6 & 4692. & 40.46 & 11.57 & 0.998 \\
 14 & 3 & 6511. & 37.31 & 12.01 & 0.861 \\
 15 & 2 & 7552. & 64.35 & 12.01 & 0.627 \\
 16 & 4 & 2020. & 46.74 & 11.60 & 0.978 \\
 17 & 2 & 5588. & 89.09 & 12.60 & 0.357 \\
 18 & 5 & 5340. & 44.84 & 11.72 & 0.987 \\
 19 & 10 & 4461. & 65.50 & 12.44 & 0.998 \\
 20 & 2 &  6200. & 20.51 & 10.49 & 0.939 \\
 21 & 2 &  4590. & 91.92 & 10.87 & 0.854 \\
 22 & 2 &  4712. & 28.28 & 11.72 & 0.841 \\
 23 & 2 &  1298. & 23.33 & 11.83 & 0.648 \\
 24 & 2 &  5102. & 46.67 & 11.63 & 0.811 \\
\end{tabular}
\end{center}
\end{table}

\end{document}